\documentclass{vgtc}                          

\ifpdf
  \pdfoutput=1\relax                   
  \usepackage{graphicx}                
  \DeclareGraphicsExtensions{.pdf,.png,.jpg,.jpeg} 
\else
  \usepackage{graphicx}                
  \DeclareGraphicsExtensions{.eps}     
\fi%

\graphicspath{{figures/}{pictures/}{images/}{./}} 

\usepackage{microtype}                 
\PassOptionsToPackage{warn}{textcomp}  
\usepackage{textcomp}                  
\usepackage{mathptmx}                  
\usepackage{times}                     
\usepackage{cite}                      
\usepackage{tabu}                      
\usepackage{booktabs}                  
\usepackage{hyperref}

\onlineid{1035}

\vgtccategory{PacificVis 2024 Posters}

\vgtcinsertpkg

\title{A Visual Analytics System to Understand \\ Behaviors of Multi Agents in Reinforcement Learning}

\author {Changhee Lee\thanks{e-mail: chlee0811@hufs.ac.kr}\\ %
     \parbox{1.4in}{\scriptsize \centering Hankuk University of Foreign Studies}%
\and Jeongmin Rhee\thanks{e-mail: 202102667@hufs.ac.kr}\\ %
    \parbox{1.4in}{\scriptsize \centering Hankuk University of Foreign Studies} %
\and DongHwa Shin\thanks{e-mail: dhshin@kw.ac.kr, corresponding author}\\ %
     \scriptsize Kwangwoon University}

\abstract{
Multi-Agent Reinforcement Learning (MARL) is a branch of machine learning in which agents interact and learn optimal policies through trial and error, addressing complex scenarios where multiple agents interact and learn in the same environment at the same time. Analyzing and understanding these complex interactions is challenging, and existing analysis methods are limited in their ability to fully reflect and interpret this complexity. To address these challenges, we provide MARLViz, a visual analytics system for visualizing and analyzing the policies and interactions of agents in MARL environments.
The system is designed to visually show the difference in behavior of agents under different environment settings and help users understand complex interaction patterns.
In this study, we analyzed agents with similar behaviors and selected scenarios to understand the interactions of the agents, which made it easier to understand the strategies of agents in MARL.
} 

\begin{document}


\firstsection{Introduction}

\maketitle


Reinforcement Learning (RL) is an integral part of machine learning in which agents learn through trial and error in a given environment, learning how to choose the optimal action to achieve their goal.
 In this process, the agent evaluates its behavior based on the reward it receives from the environment and uses it to improve its policies. In this reinforcement learning, Multi-Agent Reinforcement Learning (MARL)~\cite{hu1998multiagent} deals with situations where multiple agents are interacting and learning in the same environment at the same time.

MARL is particularly important when agents need to predict each other's actions and develop optimal policies.
For example, autonomous vehicles, robotics, and game theory, MARL is essential to understand complex interactions and derive optimal policies.
Playback visualization for episodes are one way to view agent behavior in reinforcement learning, but they have several limitations.
First, the playback visualization method can be problematic in a MARL environment. This visualization method requires viewing the agent's behavior over the course of the scenario, which takes a significant amount of time to compare the agent's behavior in a MARL environment where agents are trained with different environment settings (e.g., game mode, number of agents, reward function) and the scenario contains multiple agents.
Second, in MARL's playback visualization, all agents act simultaneously, making it difficult to distinguish and understand each agent's individual behavior and its impact on other agents or the overall system in real time.

To address these issues, we provide MARLViz, a visual analytics system to help users understand behaviors of multi agents in reinforcement learning processes.
We adopted the well-known snake game as a target problem of MARL.
MARLViz enables simultaneous comparison of the behaviors of agents trained in different environmental settings.
In addition, it also allows users to select scenarios containing agents with similar behaviors to see how they interact with each other.

\section{Task Summary}
In MARL, knowing and understanding how agents behave and affect each other is crucial. It includes comparing their behaviors and understanding their interactions, which can provide valuable insights for researchers or practitioners using RL to train agents.
To this end, we established the two core tasks that our MARLViz system should provide as follows:

\textbf{Task 1: Compare behavioral differences across multiple agents at once}
In RL, it is important to compare the behavior of agents as they are trained to achieve a goal.
For MARL in particular, there are many agents in a single scenario.
However, when visualized as playback visualization for episodes, it takes as much time as the length of the scenario to check the behavior of the agents, and it takes a considerable amount of time to compare the behavior of the agents.
To solve this issue, we used the autoencoder to learn  each agent's actions to extract feature vectors for each agent.
Then, we show those features on a dimensionality-reduced 2D scatter plot.

\textbf{Task 2: Understanding how agents interact in a scenario}
Interaction in MARL refers to the process of multiple agents acting in the same environment, influencing each other, such as cooperation, competition, and various other dynamic relationships.
In MARL, understanding the interactions of agents is essential to understand their complex dynamic relationship with the environment to design learning strategies.
MARLViz summarizes the movements of the agents, which allows users to understand the interactions between them in a simple and clear way in a short time.
This visualization can be useful for understanding the different interaction patterns that occur in MARL.

\begin{figure*}[t]
    \centering
    \includegraphics[width=1\linewidth]{figures/MARLViz_figure1.png}
    \vspace{-5mm}
    \caption{The visual interface of MARLViz. (A) The Overview shows the dimensionality-reduced 2D plot of the entire agents' features.
    (B) The Config View shows the distributions of environment settings for the game modes, the number of agents, and reward functions of the brushed agents.
    (C) In the Scenario View, each list item contains the percentage of agents' actions and reward types in the scenario trained with a specific environment setting.
     (D) In the Interaction View, the heatmap shows the agents' moving paths for the selected scenario in the Scenario View. (E) The line chart represents the trend and events of the rewards obtained by the agents in the selected scenario.
    }~\label{fig:user_interface}
    \vspace{-7mm}
\end{figure*}

\begin{figure}[b!]
    \centering
    \vspace{-2mm}
    \includegraphics[width=1\linewidth]{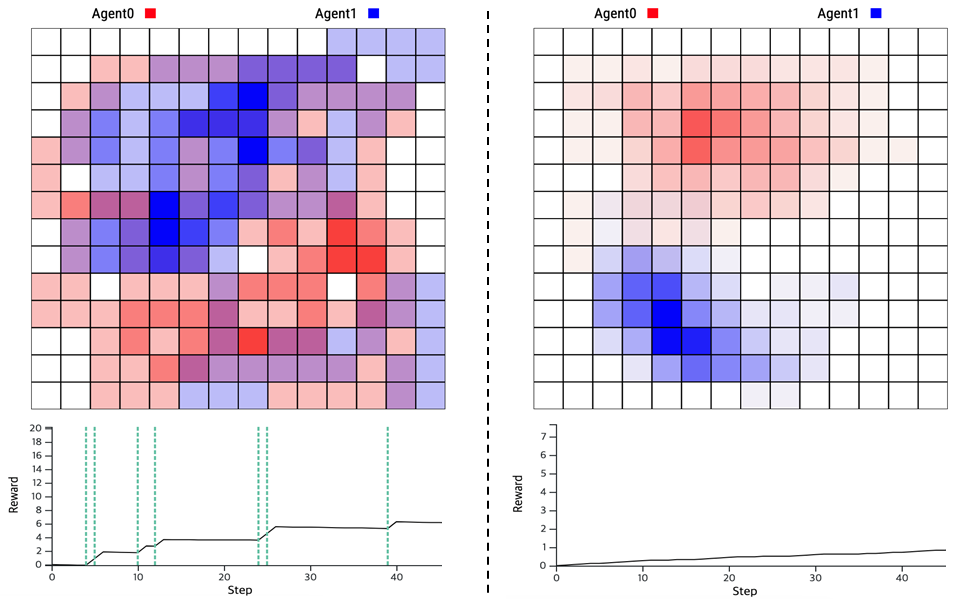}
    \vspace{-6mm}
    \caption{Visualization of the agents interactions in Scenario View. On the left is a case with a high proportion of "straight" behaviors, high "fruit" rewards, and low "time" rewards, and on the right is a case with a low proportion of "straight" behaviors, low "fruit" rewards, and high "time" rewards.}~\label{fig:scenario}
    \vspace{-8mm}
\end{figure}

\section{Visual Interface}
A visual interface of MARLViz (\autoref{fig:user_interface}) consists of four main views: Overview, Config View, Scenario View, and Interaction View, each of which provides users with different analysis capabilities.
By utilizing these views on a single screen simultaneously, users can start with a summary of the agents and drill down into the details of each scenario.

An \textbf{Overview} (\autoref{fig:user_interface}-A) is the entrance to MARLViz, providing a summarized view of all agents. This view provides a scatter plot of the similarity of agents' step-by-step actions and events.
Agents with similar action patterns appear clustered, and users can use brushes to filter the clusters or specific agents of interest.
By comparing the environment settings features of differently clustered agents in a scatter plot in Config View, users can compare differences in behavior across multiple agents at once.

A \textbf{Config View} (\autoref{fig:user_interface}-B) provides a visual representation of the distribution of environment settings of the agents brushed in the Overview.
The settings include \textit{game modes}, \textit{the number of agents}, and \textit{reward functions} (\textit{time} and \textit{death}).
We represent the game modes and the number of agents in pie charts because it is important for users to perceive the difference in their proportions.
The reward functions consists of two variables, and since we need to represent the frequency of each, we depicted it using a heatmap.
The heatmap shows the most frequently set reward values in dark color for the brushed group of agents.
It allows users to identify common environment settings among the brushed agents.

A \textbf{Scenario View} (\autoref{fig:user_interface}-C) is where users can analyze and compare the action behaviors of the agents brushed in the Overview in each scenario.
We visualize each scenario as a list item.
Each list item contains the environment setting, action rates, and reward distributions.
We visualize the action rates as a pie chart, and the reward distributions as a bar chart.
It allows users to analyze the pattern of actions taken by the agents in each scenario and the types and amounts of rewards they earned.

An \textbf{Interaction View} (Fig. \ref{fig:user_interface}-D and \ref{fig:user_interface}-E) provides a detailed information of the interactions of the agents within a specific scenario selected in the Scenario View.
It contains a heatmap and a line chart.
The heatmap displays the agents' moving paths in their game.
Each agent appears as a color with a different hue, and the saturation of each color is proportional to the number of times the agent has visited that cell.
The line chart shows rewards earned and events over time, showing the interactions of the agents based on their action behaviors.
The vertical dotted lines, which are a visual representation of when events occur, are colored the same as the bar charts that represent reward values in the Scenario View.

\section{Usage Scenario}



We analyzed 216 agents from 72 different scenarios using MARLViz.
All agents are represented as points in the Overview. In the scatterplot, clustered points indicate agents with similar behaviors.
After seeing the distribution of the scatterplot, we brushed a agent cluster to see the distributions of their environment settings in the Config View.
We found that clustered agents were concentrated in specific environment settings.
It indicates that the environment settings strongly correlate with the agent behaviors (Task 1).

We also analyzed the scenarios that the agents belong to in the Scenario View and the Interaction View.
In the Scenario View, there were two notable patterns.
one with a higher proportion of "straight" than other behaviors, with high "fruit" rewards and low "time" rewards, and the other with a lower proportion of "straight" than other behaviors, with low "fruit" rewards and high "time" rewards.
We also noticed two opposing patterns in the Interaction View. (\autoref{fig:scenario}) 
In the first case, the agents feed more frequently and use a larger area of the map.
On the other hand, in the second case, the agents feed less frequently and use a smaller map.

\section{Conclusion}
MARLViz helps to understand MARL by allowing users to compare and analyze the behavior of agents in MARL, and to easily understand the interactions of agents by selecting the scenarios in which the agents they want to check belong to.
In future research, we aim to provide visualizations and interactions that can be generalized beyond the snake game to various MARL environments.


\bibliographystyle{abbrv-doi}

\bibliography{reference}

@inproceedings{hu1998multiagent,
  title={Multiagent reinforcement learning: theoretical framework and an algorithm.},
  author={Hu, Junling and Wellman, Michael P and others},
  booktitle={ICML},
  volume={98},
  pages={242--250},
  year={1998}
}
\end{document}